\documentclass[superscriptaddress,twocolumn,prb,showpacs]{revtex4}
\usepackage[dvips]{graphicx}
\usepackage{amsmath,amsfonts,amssymb,bm,ulem}
 \newcommand{\m}{{\text{M}}}
 \newcommand{\mi}{{\text{m}}}
 \usepackage[usenames]{color}

  \newcommand{\br}{\mathbf{r}}

 \newcommand{\cL}{\mathcal{L}} \newcommand{\cN}{\mathcal{N}}

\begin{document}
\title{Slow relaxation of conductance of amorphous hopping insulators}
\author{A. L. Burin}
\affiliation{Department of Chemistry, Tulane University, New
Orleans, LA 70118, USA} \affiliation{Pacific Institute of Theoretical Physics,
University of British Columbia,
6224 Agricultural Road,
Vancouver, BC
Canada, V6T 1Z1}
\author{V. I. Kozub}
\affiliation{A. F. Ioffe  Physico-Technical Institute of Russian
Academy of Sciences, 194021 St. Petersburg, Russia}
\affiliation{Argonne National Laboratory, 9700 S. Cass Av.,
 Argonne, IL 60439, USA}
\author{Y. M. Galperin}
\affiliation{Department of Physics and Center for Advanced materials \&
  Nanotechnology, University of Oslo, PO Box
1048 Blindern, 0316 Oslo, Norway} \affiliation{A. F. Ioffe
Physico-Technical Institute of Russian Academy of Sciences, 194021
St. Petersburg, Russia} \affiliation{Argonne National Laboratory,
9700 S. Cass Av., Argonne, IL 60439, USA}
\author{V. Vinokur}
\affiliation{Argonne National Laboratory, 9700 S. Cass Av.,
Argonne, IL 60439, USA}
\date{\today}
\begin{abstract}
We discuss memory effects in the conductance of hopping insulators
due to slow rearrangements of structural defects   leading to
formation of polarons  close to the electron hopping states. An
abrupt change in the gate voltage and corresponding shift of the
chemical potential change populations of the hopping sites, which
then slowly relax due to  rearrangements of structural defects
reducing the density of states. As a
result, the density of hopping  states becomes time dependent on a
scale relevant to rearrangement of  the structural defects leading
to the excess time dependent conductivity.
\end{abstract}

\pacs{73.23.-b 72.70.+m 71.55.Jv 73.61.Jc 73.50.-h 73.50.Td}
\maketitle

\section{Introduction}

A peculiar transport memory effect has been observed in many hopping
insulators~\cite{Chorin93,Martinez97,Ovadyahu97,Grenet03,Grenet07}:
Shifting the system either from  the equilibrium or from its steady
transport state produced by,  e.g., a sudden change of the gate
voltage, $\delta V_g$, increases the conductance $\sigma$ of the
system, the effect not depending on the sign of the change.  The
$\sigma(\delta V_g)$ dependence shows a characteristic memory cusp
(see Ref.~\onlinecite{Ovadyahu06} for experimental details and a
review) which  may persist for a long time.  Explanations of this
appealing observation can be grouped into two major classes which are
often referred to as \textit{intrinsic} and \textit{extrinsic}
mechanisms. The former one attributes memory effects to slow dynamics
of strongly correlated electrons subject to quenched disorder and thus
forming a \textit{Coulomb glass}~\cite{Vaknin02,Vaknin98,Clare}.  The
connection between the features of a glassy phase and the formation of
a Coulomb gap was revealed in Ref.~\onlinecite{Muller04}, suggesting
that within the locator approximation the correlated electronic system
maps to the Sherrington-Kirkpatrick spin glass. A direct relation
between the long range electron correlations and the formation of the
exponential distribution of deep electron states which is
characteristic to glassy systems was also demonstrated
in Ref. \onlinecite{GVG07}.  In a Coulomb glass, memory effects reflect delayed
formation of the Coulomb gap in the Efros-Shklovskii (ES) regime of
the variable range hopping (VRH).\cite{Mueller05} Recently we have
demonstrated how ``slow'' many-electron fluctuators  can be formed in
a Coulomb system and analyzed their influence on the $1/f$
noise.\cite{Burin} Such fluctuators can also lead to slow relaxation
and memory effects.

An extrinsic scenario  assumes that electronic memory effects are
caused by slowly relaxing \textit{atomic} configurations influencing
conducting channels and was first proposed in
Ref.~\onlinecite{Adkins84} to explain the $G(V_g)$-cusp in granular Au
films.  A possibility that polaron effects may be responsible for slow
relaxation in hopping conductors was also discussed in
Refs. \onlinecite{Vaknin00,Grenet07}.

The Coulomb glass is formed  well below the temperature $T_{0}$
entering the ES law for the VRH as  
\begin{equation}
\sigma=\sigma_{0}e^{-(T_0/T)^{1/2}}.
\label{eq:ES}
\end{equation}
Experiment~\cite{McCammon} has shown that in this regime the $1/f$
noise intensity is strongly correlated with $T_0$ proving electronic
nature of the low-frequency noise. However, the conductance memory
cusps and their relaxation (ageing) where observed in the systems
which do not exhibit ES behavior, but are apparently subject to
structural disorder~\cite{Ovadyahu06,Ovadyahu07}. Various memory
effects were observed in metallic granular
structures~\cite{Adkins84,Martinez97,Grenet07}, which also possess a
high degree of structural disorder. This calls for careful examining, whether
the memory effects and the ``two-cusps"
$G(V_g)$-dependence~\cite{Ovadyahu06} can be explained as a result of
slow relaxation in the structurally disordered atomic matrix.

 In this Letter we present a model that may serve as a step toward a
quantitative description of memory cusps based on an extrinsic
mechanism.  We show that due to slow relaxation of atomic structure,
polaron clouds, which form near the hopping sites, suppress the bare
electron density of states (DOS). Changing gate voltage shifts
chemical potential, removes the polaron screening, and, thus,
increases the hopping conductivity. As the atomic structure adjusts
itself with time to the new position of chemical potential, the
conductivity relaxes to its quasi-stationary magnitude.

Atomic structural relaxation is described by two level systems (TLS).\cite{AHVP,Hunklinger} The TLS model successfully describes thermodynamics and kinetics of amorphous solids at low temperature. It suggests that there exist atoms or groups of atoms undergoing tunneling motion and characterized by the broad universal distribution of their parameters. Since all materials where the memory effects in conductivity were observed are strongly disordered one would expect that two level systems (TLS) should exist there similarly to other glasses and disordered materials.\cite{Hunklinger}  These two-level systems interact with conducting electrons because they possess the dipole moment. In this manuscript we examine the effect of electron-TLS interaction on the non-equilibrium conductivity. We show that this interaction results in the non-equilibrium behavior of conductivity which is qualitatively equivalent to the experimental observations, i. e. increase in conductivity after gate voltage application with its subsequent logarithmic relaxation to the equilibrium value. This theory uses the previous work,\cite{Burin95,O2} where the similar non-equilibrium behavior of the dielectric constant in amorphous solids\cite{O1,O2} has been explained using the TLS interaction.
Since TLS parameters are quite universal from material to material\cite{Hunklinger} we can use these parameters for quantitative estimates which show that our theoretical predictions are consistent with existing experimental data. 

The paper is organized as following. In Sec. \ref{sec:elelint} we accurately define the conditions where our consideration is applicable, i. e. electron-electron interaction can be neglected, while electron-TLS interaction is significant. In Sec. \ref{sec:noneqDOS} the non-equilibrium behavior of electron density of states caused by their interaction with TLS is derived. In Sec. \ref{sec:noneqcond} the non-equilibrium behavior of conductivity is described and compared with the experimental data.  The results of the manuscript are summarized in conclusive Sec. \ref{sec:concl}.

\section{When the electron-electron interaction can be neglected?}
\label{sec:elelint}

In this manuscript we ignore the effect of electron-electron interaction on the non-equilibrium behavior of conductivity.  This is possible only under specific conditions when the electronic interaction is weak compared to their characteristic energies and there is no slow relaxation within the electronic subsystem. According to various considerations \cite{Mueller05,Burin} the slow relaxation in electronic subsystem can take place only below some critical temperature $T_{G}$ which is defined as the electronic glass transition temperature in Ref. \onlinecite{Mueller05}. This transition temperature is defined by the Coulomb gap energy 
\begin{equation}
k_{B}T_{G} \sim \Delta_{C}.  
\label{eq:tg}
\end{equation}
Below we set $k_{B}=1$. 

The Coulomb gap depends on the electronic density of states $g_{0}$, dielectric constant $\kappa$, system dimension $d$and electron localization radius $a$. We believe that the dimensionless parameter $\chi=g_{0}a^{d-1}e^{2}/\kappa=g_{0}a^{d}T_{0}$ is small, i. e. 
\begin{equation}
\chi = g_{0}a^{d}T_{0}<1,  
\label{eq:smallchi}
\end{equation}
where the characteristic temperature $T_{0}$ in the Efros-Shlovskii hopping conductivity Eq. (\ref{eq:ES}) is given by   
\begin{eqnarray}
k_{B}T_{0} \sim \frac{e^{2}}{\kappa a}.    
\label{eq:TES}
\end{eqnarray}
Then the Coulomb gap is defined as 
\begin{equation}
\Delta_{C}= T_{0}\chi^{\frac{1}{d-1}}.  
\label{eq:CoulGap}
\end{equation}
If temperature exceeds the Coulomb gap energy then electronic relaxation is fast.  It is also useful to notice  that for the electronic excitations with typical energy $E$ their characteristic interaction $U_{E}\approx e^{2}(g_{0}E)^{\frac{1}{d}}$ is smaller than the energy $E$ until $E > \Delta_{C}$ and the condition $U_{E}\approx E$ serves as the definition of the Coulomb gap. So when $T>\Delta_{C}$ the interaction of representative excitations with energy $E \geq T$ is less than their energy so it can be treated as weak.  

If the system is near metal dielectric transition\cite{McCammon} and the electron localization radius is large, then one can possibly have the opposite limit $\chi = g_{0}a^{d-1}e^{2}/\kappa > 1$,   This is the case which takes place in silicon Mosfets investigated in Ref. \onlinecite{McCammon}. Under those conditions the only Efros-Shklovskii variable range hopping law Eq. (\ref{eq:ES}) is observed. 
We believe this is not the case for the systems of interest,\cite{Chorin93,Martinez97,Ovadyahu97,Grenet03,Grenet07} where the different conductivity behavior is observed more close to the Mott's variable-range hopping law, which suggests $\chi<1$.\cite{ESbook} 

Therefore it is not possible to interpret the memory effects in conductivity at $K_{B}T > \Delta_{C}$ without the involvement of electronic interaction with the extrinsic slowly relaxing defects like TLS. Our theoretical study is restricted to this ``high temperature'' situation. Since the temperature $T_{0}$ can change within the range of $1-100$K \cite{McCammon} our assumption that the experimental temperature $4$K exceeds the Coulomb gap energy Eq. (\ref{eq:CoulGap}) does not conflict with the common sense.  We believe that at least some of experiments \cite{Chorin93,Martinez97,Ovadyahu97,Grenet03,Grenet07} are performed in this temperature range. 
Unfortunately, existing experimental data do not permit us to answer the question whether experimental temperature is above or below the Coulomb gap. Although the absence of the Efros-Shklovskii conductivity temperature dependence Eq. (\ref{eq:ES}) \cite{Ovadyahu97} agrees with our assumption that the electronic interaction is not important this does not prove the absence of Coulomb gap and electronic glassy state.\cite{ESbook}. We therefore suggest additional experiments which can help to investigate the Coulomb gap in the system which is the necessary prerequisite of the Coulomb glass state. Such experiment can be made for instance using scanning tunneling microscopy.\cite{Adams} If the Coulomb gap in the density of electronic states will not be observed at experimental temperatures then one should expect that the extrinsic mechanism is in charge for the memory behavior in conductivity. 

\section{Electron density of states affected by electron-TLS interaction}
\label{sec:noneqDOS} 

Our consideration is based on the concept of `two level systems'
(TLS)~\cite{AHVP} taking as its central hypothesis the assumption
that in a system with quenched disorder a certain number of atoms
(or groups of atoms) can occupy one of (at least) two equilibrium
positions.  These atoms therefore move in a double-well potential
created by their environment and characterized by the asymmetry
energy (difference in energy minima) and by the height and the width
of separating barrier.  The atoms comprising the TLS change their
configuration  either by tunneling through- or by thermally
activated hopping over the barrier. Randomness in the heights and
widths of the TLS barriers gives rise to exponentially broad
distribution of structural relaxation times.
\begin{figure}[b]
\centering
\includegraphics[width=8cm]{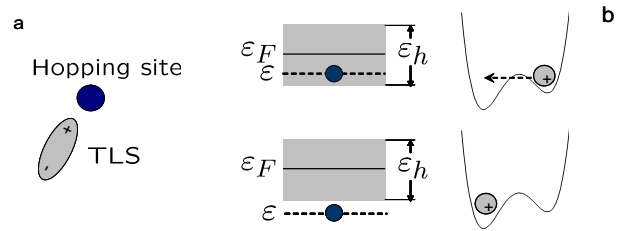}
\caption{a -- Polaron formed by an electron site and an
    adjacent TLS;
 b -- Transitions of the TLS between its states shifts
  the electron energy with respect to the Fermi level and can bring it
  outside the hopping band having the width $\varepsilon_h$.
  \label{fig:polaron} }
\end{figure}

A localized electric charge polarizes the neighboring TLSs, which
thus acquire the electric dipole moment, $\bm{\mu}$, and screen the
original bare charge, see Fig.~\ref{fig:polaron}a. The interaction
energy of TLS with the localized charge is estimated as
\begin{equation} \label{eq:u1}
U(\br)\sim e(\bm{\mu}\cdot \br)/\kappa r^3\, , \quad r \gg a \, ,
\end{equation}
where $\br$ is the vector connecting the center of this TLS with the
position of localized electron state, $a$ is the localization
length, and $\kappa$ is the material dielectric constant.  Polarized
TLSs form a polaron cloud around the localized electron and create a
polaron gap for electronic excitations, implying that the states
with the single-electron energy $\phi < \sum_i U (\br_i)$ cannot be
excited any more.

In what follows we will find the correction to the electronic DOS,
$g_0$, due to electron-TLS interactions in the lowest approximation
in the TLSs density. To this end we first determine the change in
DOS due to a single neighboring TLS characterized by the given
energy splitting, $E$, and the relaxation time, $\tau$. The target
correction to the electron DOS is then obtained by averaging of the
single TLS contribution with respect to all possible neighboring TLS
positions and $E$ and $\tau$.

Consider the correction to the density of electronic states with some energy $\varepsilon$ caused by electron-TLS interaction.
Since the concentration of TLS is small we can assume that only one most closely located two-level system is important, while the probability that two TLS are significant is much smaller than unity.  Then we consider electron interacting with the single TLS using the standard Hamiltonian 
\begin{equation}
\widehat{H}_{pair}=\phi n + Un\sigma^{z} +\Delta \sigma^{z}, 
\label{eq:HamPair}
\end{equation}
where $n=0, 1$ is the electron population operator, the spin $1/2$ $z$-projection operator $\sigma^{z}=\pm 1/2$ describes two states of the TLS, $\phi$ is the electronic energy in some localized state, $\Delta$ is the energy of the two level system and $U$ is the charge-dipole interaction of the electron with the TLS. In the thermal equilibrium the electron excitation energy $\varepsilon=\phi + U\sigma$ can take values $\phi\pm U/2$ with the probabilities defined by the equilibrium Boltzmann factors  
\begin{widetext}
\begin{eqnarray}
P^{+}=\frac{\exp\left(-\frac{\Delta}{2T}\right)+\exp\left(\frac{-\Delta/2-U/2-\phi}{T}\right)}{\exp\left(\frac{-\Delta}{2T}\right)+\exp\left(\frac{-\Delta/2-U/2-\phi}{T}\right)+\exp\left(\frac{\Delta}{2T}\right)+\exp\left(\frac{\Delta/2+U/2-\phi}{T}\right)},  
\nonumber\\ 
P^{-}=\frac{\exp\left(\frac{\Delta}{2T}\right)+\exp\left(\frac{\Delta/2+U/2-\phi}{T}\right)}{\exp\left(\frac{-\Delta}{2T}\right)+\exp\left(\frac{-\Delta/2-U/2-\phi}{T}\right)+\exp\left(\frac{\Delta}{2T}\right)+\exp\left(\frac{\Delta/2+U/2-\phi}{T}\right)},  
\label{eq:probplmin}
\end{eqnarray}
\end{widetext}
respectively.  The time-dependent correction to the density of states taken at the certain time $t$ can be associated only with those two level systems which have the relaxation time $\tau$ longer than the time of the experiment $t$.  That time must also have upper restriction by some maximum TLS relaxation time $\tau_{max}$ which serves as the upper cutoff in the TLS logarithmically uniform distribution over their relaxation times $P(\Delta, \tau)=P_{0}\Theta(\tau_{max}-\tau)/\tau$. 
Within the logarithmic accuracy one can account for the above constraints introducing integrated time dependent TLS density as 
\begin{equation}
P(t)=P_{0}\ln(t/\tau_{\mi}).
\label{eq:modif_DOS}
\end{equation} 
This expression   clearly demonstrates the nature of logarithmic time dependence of TLS contribution to the electronic of states similarly to previous work \cite{Burin95}. 

The correction  to the density of states of the particular electron with energy $\varepsilon$ caused by its interaction with neighboring TLSs can be expressed as
\begin{widetext}
\begin{equation}
\delta g_{s}(\varepsilon) = \frac{1}{V}\sum_{ij}\left(P_{ij}^{+}\delta(\varepsilon-\phi_{i}-U_{ij}/2)+P_{ij}^{-}\delta(\varepsilon-\phi_{i}+U_{ij}/2)-\delta(\varepsilon-\phi_{i})\right), 
\label{eq:dosTLS_correction}
\end{equation}
\end{widetext}
where the sum is taken over all pairs made of an electron $i$ and a TLS $j$ characterized by energies $\phi_{i}$ and $\Delta_{j}$, respectively, $U_{ij}$ is their interaction and $V$ is the system volume. Probabilities $P_{ij}^{\pm}$ are defined using Eq. (\ref{eq:probplmin}) with the substitution $\phi=\phi_{i}$, $\Delta=\Delta_{j}$ and $U=U_{ij}$.  
The summation in Eq. (\ref{eq:dosTLS_correction}) can be replaced with the integration over electron and TLS energies and the distance $R$ between electron and TLS. This yields  
\begin{widetext}
\begin{eqnarray}
<\delta g(\varepsilon)> = g_{0}P(t)\int d^{d}{\bf R} \int_{-\infty}^{\infty} d\phi \int_{-\infty}^{\infty} d\Delta 
\nonumber\\ 
\times \left<\frac{(1+e^{-\frac{\varepsilon}{T}})\delta(\varepsilon-\phi-U/2)}{1+e^{-\frac{\varepsilon}{T}}+e^{\frac{\Delta}{T}}(1+e^{-\frac{\varepsilon-U}{T}})}+\frac{(1+e^{-\frac{\varepsilon}{T}})\delta(\varepsilon-\phi+U/2)}{1+e^{-\frac{\varepsilon}{T}}+e^{\frac{-\Delta}{T}}(1+e^{-\frac{\varepsilon+U}{T}})}-\delta(\varepsilon-\phi)\right>,  
\label{eq:dosTLS_correction_simpl}
\end{eqnarray}
\end{widetext}
where $g_{0}$ is the electronic density of states at energy $\varepsilon$ taken in the absence of interaction with TLS and $d$ is the system dimension. Remember that $U=e({\bf \mu, R})/(\kappa R^3)$ is the dipole charge interaction between electron and TLS possessing the dipole moment ${\bf \mu}$ and averaging is performed with respect to random directions of a TLS dipole moment. After integration over electronic energy $\phi$ and TLS energy $\Delta$ in Eq. (\ref{eq:dosTLS_correction_simpl}) we obtained 
\begin{eqnarray}
<\delta g(\varepsilon)> = g_{0}P(t)T
\nonumber\\
\times\int d^{d}{\bf R}\left<\ln\left(\frac{\cosh(\varepsilon/T)+1}{\cosh(\varepsilon/T)+\cosh(U/T)}\right)\right>. 
\label{eq:dosTLS_correction_step2}
\end{eqnarray}
The most important electrons are those contributing to the hopping conductivity. These electrons have energy $\varepsilon$ of order of the hopping energy $\varepsilon_{h}$ which always exceeds the thermal energy. Therefore assuming $\varepsilon>T$ one can approximate the logarithm under the integral in leading order in $1/T$ as  $\ln\left(\frac{\cosh(\varepsilon/T)+1}{\cosh(\varepsilon/T)+\cosh(U/T)}\right)\approx -\frac{||U|-\varepsilon|}{T}\theta(|U|-\varepsilon)$. 
The final expression for the correction is similar to the one used in earlier works \cite{Baranovskii80,Burin95} for the Coulomb gap and the dipole gap in the density of TLS states. However our derivation of the correction is more general because our expression Eq. (\ref{eq:dosTLS_correction_step2}) can be used at arbitrarily temperature while the earlier derivation is valid only in the low temperature limit. Finally the correction to electronic density of states can be expressed as  
\begin{eqnarray}\nonumber
\delta g(\varepsilon,t) &\approx& -2g_{0}P_0 \cL \int d\br\
[U(\br)-\varepsilon]\, \theta[U(\br)-\varepsilon]\nonumber\\
&\approx&   - \frac{8\pi}{3}
\left(\frac{e\bar\mu}{\kappa}\right)^{3/2}
\frac{g_0P_0}{\sqrt\varepsilon} \ln\left(\frac{t}{\tau_\mi}\right)\,
, \label{eq:ans3}
\end{eqnarray}
 where $\bar{\mu}$ is the typical dipole moment of a TLS and $\tau_\mi$ is some characteristic minimum time associated with the gate voltage application.
Only the electrons with the energy $\varepsilon \sim \varepsilon_{h}=T^{3/4}a^{-3/4}g^{-1/4}$ influence the hopping
conductance.  We believe that at $\varepsilon \sim \varepsilon_{h} \sim 30$K Eq. (\ref{eq:ans3}) is still applicable.   

\section{Analysis of experimental data}
\label{sec:noneqcond}

Equation (\ref{eq:ans3}) can be used to interpret the experimental
data only if all relevant TLSs were initially out of equilibrium. In
other words, all electrons contributing to hopping must be surrounded
by non-equilibrium TLSs. The equilibrium can be broken due to the
application of the gate voltage. This can be realized when all
electrons contributing to hopping are ``newcomers'', i.e., they are
brought to the vicinity of the Fermi level by application of the gate
voltage, $V_{g}$. For that the shift of the Fermi energy by the gate
voltage must exceed the hopping energy $\varepsilon_{h}$. Then those
electrons break the equilibrium in their neighboring TLS randomly
changing their energies by the scale of their interaction with those
TLS. All relevant TLS with the energy of order of $\epsilon_{h}$
coupled to entering electrons (or holes) by the interaction having the
same order of magnitude experience the jump in their energy induced by
entering electrons. This moves them all out of equilibrium. Their
relaxation leads to the polaron shift of electron energy out of the
Fermi energy thus reducing the conductivity. Thus the condition
$\Delta E_{F} \approx \varepsilon_{h}$ defines the  width of the cusp
in the non-equilibrium conductivity as a function of the gate
voltage. One can show using Ref.~\onlinecite{Vaknin02} that at those
conditions the shift of TLS energy induced ``directly'' by the gate
voltage is still smaller than $\epsilon_{h}$ because of the small TLS
dipole moment $\bar{\mu} \approx 2\cdot 10^{-18}$
erg$^{1/2}$cm$^{3/2}$ \cite{Hunklinger}. We can compare this estimate
with the experiment extracting $\epsilon_{h}$ from the expression for
the conductivity, $\sigma \sim
\sigma_{0}\exp(-\varepsilon_{h}/T)$. Let us put $\sigma_0$
equal to the minimal metallic conductivity ($\sim 10^{-4}$ Ohm$^{-1}$)
and use available experimental results of Ref.~\onlinecite{Ovadyahu97} for
the representative sample with resistance of $3.8$ MOhm  at $T=4.1$ K. 
The assumption about $\sigma_{0}$ is justified by the experimental observations 
(see Ref. \cite{PS} and references therein). 

Then  the hopping energy is
$\varepsilon_{h}=T\ln(\sigma_{0}/\sigma)\approx 2.1\cdot
10^{-3}$ eV. This value agrees qualitatively with the shift of the
Fermi energy $3-5$ meV associated with the value of the gate voltage
$V_{g}$   at which the memory cusp is affected as was estimated
earlier in Ref.~\onlinecite{Vaknin02}. Thus the suggested mechanism agrees
with the experimental observations.

The correction to the conductivity can be estimated as
$[(\varepsilon/T)\delta g(
\varepsilon, t)/g_0]_{\varepsilon=\varepsilon_h}$. In this way we get
\begin{equation}\label{G}
\frac{\delta \sigma (\delta V_g,t) }{\sigma}
\simeq \frac{P_0 e \bar{\mu}}{\kappa}\
  \left[\frac{e\bar{\mu}}{\kappa \varepsilon_h }\right]^{1/2}
  \ln \left(\frac{t}{\tau_\mi}\right)
  \ln\frac{\sigma_0}{\sigma(T)}.
\end{equation}
The time-dependent factor $\cL\equiv \ln
(t/\tau_\mi)$ contains the
measurement time as $t$ and the inverse sweep rate, $\tau_{\mi} \gg
\tau_h=\tau_0 e^{(T_\m/T)^{1/4}}$, and affects only
the amplitude of the peak, but not its shape. Here $\tau_{h}$  is the characteristic time of the variable range hopping and the preexponential factor $\tau_{0}$ is of order of $1$ps at experimental temperatures. This can explain the
experimentally observed independence of the dip shape on the sweep
rate~\cite{Vaknin02}.

Similar considerations apply to the ES VRH with the proper
renormalization of the hopping parameters. Now  the typical energy
scale  optimizing the hopping rate is $\varepsilon_h =(T_0T)^{1/2}$,
where $T_0=\beta e^2/\kappa a$, $\tau_h=\tau_0e^{(T_0/T)^{1/2}}$,
and $\beta \approx 2.8$ is a numerical factor~\cite{ESbook}.

Let us now discuss the available experimental data in light of the
above theory.  According to our previous estimates, $\varepsilon_{h}
\approx 2.1\cdot  10^{-3}$ eV. The typical dipole moment can be
estimated as $\bar{\mu} \sim 2\cdot 10^{-18}$  erg$^{1/2}$cm$^{3/2}$,
which is close to typical dipole moments of TLSs in glasses
\cite{Hunklinger,Burin95}.

We are not aware of independent measurements of the TLS density, $P_0$, in the
materials under consideration. In principle, this quantity can be
determined, e.\,g., by measurement of the low frequency dielectric
constant at $T\lesssim 1$~K, where it should depend on the temperature
logarithmically, the
slope of the logarithmic dependence being $P_{0}\mu^{2}/\kappa$.
This dimensionless quantity turns out to be almost the same, $\approx
0.3\cdot 10^{-3}$,  in many materials with strong quenched disorder, see
\cite{Hunklinger,Burin95,O1,O2} and references therein. Apparent
universality of this quantity in such materials was attributed to the
interaction between TLSs~\cite{O2,Yu2}. Assuming
that the material studied in \cite{Ovadyahu97} belongs to the same
``universality class" as the materials with strong quenched disorder
we estimate $P_0$ as $0.6\cdot 10^{33}$ erg$^{-1}$cm$^{-3}$, as
in the most oxide glasses~\cite{Hunklinger}. We set $\kappa \sim 10$ following Ref.~\onlinecite{Clare}.

Now we can check our assumptions regarding 3D arrangement of TLSs
forming the polarons and regarding lowest approximation in the TLS
density. Estimating the polaron radius, $\bar{r}$, as as the
length at which the electron-TLS interaction, $e\mu/\kappa
\bar{r}^2$ is comparable with the typical electron energy
$\varepsilon_{h}$ we get $\bar{r} \approx 1$ nm that is much less than
the sample thickness $\ell$. The average number of the TLS forming
polarons, i.\,e., located within the polaron radius and having $E\lesssim
\varepsilon_h$,
$\cN \sim (4\pi/3) P_{0}\varepsilon_{h}\bar{r}^{3}$ turns
out to be $\sim 10^{-2}$, i.\,e., much less than $1$. Thus the lowest
approximation in the TLS density is valid.

Using the above estimates and Eq.~(\ref{G}) we get  $\delta \sigma (t)
/\sigma  \approx 0.02 \ln(t/\tau_\mi)$. Thus we predict the
logarithmic relaxation rate of conductivity  $r=d\ln(\sigma)/d \ln(t)
\approx 0.02$. According to the  experimental data
\cite{Ovadyahu97,Vaknin98}  at $T=4.1$K and for the sample thickness
equal to $10$nm the conductivity changes by about $8\%$ during two
decades in time so we can estimate it as $r\approx 0.015$. Thus our
theory agrees with experimental data reasonably well.

\section{Discussion and conclusion.} 
\label{sec:concl}

We have presented a simple model of slow dynamics of hopping
conductance in structurally disordered hopping insulators.  It takes
into account rearrangements of the dynamic structural defects, TLSs,
leading to formation of polarons close to the electron hopping
states. The model qualitatively explains both the logarithmic
relaxation and memory effects, and provides quantitatively reasonable
estimates of the time-dependent non-equilibrium change in
conductivity, $\delta \sigma(V_g,t)$ (see, e.\,g., \cite{Ovadyahu06},
Fig. 2).

The dependencies of this quantity on different parameters -- electron
concentration, controlled by the gate voltage; magnetic
field,~\cite{Ovadyahu97}, and  various protocols of breaking down the
system equilibrium -- are encoded in the logarithmic factors
$\ln(\sigma/\sigma_{0})$ and $\ln (t/\tau_\mi)$ while the temperature
dependence enters as a power law through the energy $\varepsilon_h$:
$\delta \sigma \propto \varepsilon_h^{-3/2}$. That leads to the main
temperature dependence $\propto T^{-9/8}$ and $\propto T^{-3/4}$ for
the Mott and ES VRH, respectively.  Thus our theory explains the fast
increase of the non-equilibrium raise of conductivity with decreasing
the temperature.

The following note is in order.  We demonstrated that ``slow"
excitations induced by structural disorder can indeed be  responsible
for the double-dip memory effect in hopping semiconductors in analogy
to those in glasses~\cite{O1,O2}. The effect we considered is due to
polarons formed by the structural excitations. Yet, one should bear in
mind that a similar polaron effect appears in pure electronic
models. In particular, the polarons formed from pair excitations were
considered in \cite{polaron}. In its turn, ``electronic polarons" can
be formed from slow relaxing electronic ``aggregates'' discussed
in Ref. \onlinecite{Burin}.  The presence of structural TLS in InO films is
indirectly supported by the fact that the low-frequency noise in these
materials does not increase significantly under the disorder-driven
metal-to-insulator transition.\cite{cohen} 

Further experimental verification is necessary to decide whether the
non-equilibrium behavior is associated with the extrinsic or intrinsic
model. A regular approach to attain this goal should be based on
techniques affecting differently the structural and electronic degrees
of freedom. One of the possibilities is measurement of the AC linear
response such as simultaneous measurements of the attenuation and
velocity of the acoustic waves in transverse magnetic field. The
expected effect is caused by  transitions in structural two-well
configurations, while the electronic transitions are strongly
suppressed by the magnetic field \cite{Drichko}. Alternative experimental verification can be performed using canning tunneling microscopy\cite{Adams} as described in Sec. \ref{sec:elelint}. 

\acknowledgements
We thank Boris Shklovskii, Marcus M\"uller, Tierry Grenet,
and Michael Pollak 
for useful discussions and suggestions. AB also acknowledges Philip Stamp, Douglas Osheroff, Clare Yu and other participants of the International workshop "Mechanical Behaviour of Glassy Materials" (Vancouver, Canada, July 2007) for useful comments and stimulating discussions.  

The work of AB is supported by the Louisiana Board of
Regents (Contract No. LEQSF (2005-08)-RD-A-29). The work of YG, VK and
VV was supported by the U. S. Department of Energy Office of Science
through contract No. DE-AC02-06CH11357 and by the Norwegian Research
Council through the Norway-USA bilateral program.

\end{document}